\documentclass[letter]{aa}  

\usepackage{graphicx}
\usepackage{txfonts}

\def\kms{km\,s$^{-1}$}

\def\msun{M$_{\odot}$}

\def\rsun{R$_{\odot}$}
\def\Rsun{R$_{\odot}$}

\def\Msun{M$_\odot$}

\def\Lsun{L$_\odot$}

\def\l{$\lambda$}

\def\pbeta{Pa~$\beta$}
\def\pdelt{Pa~$\delta$}
\def\pgam{Pa~$\gamma$}
\def\peps{Pa~$\epsilon$}
\def\halp{H~$\alpha$}

\def\hbet{H~$\beta$}

\def\hgam{H~$\gamma$}

\def\hea{He\,{\sc i}}
\def\heb{He\,{\sc ii}}
\def\nc{N\,{\sc iii}}
\def\fea{Fe\,{\sc i}}

\def\mgb{Mg\,{\sc ii}}

\def\sid{Si\,{\sc iv}}

\def\xsh{X-shooter}
\def\srv{$\sigma_\mathrm{1D}$}
\def\fbin{$f_\mathrm{bin}$}

\def\pcutoff{$P_\mathrm{cutoff}$}
\defcitealias{Maria16}{Paper~I}

\titlerunning{Binarity in M17}
\authorrunning{Sana et al.}

\begin{document}

   \title{A dearth of short-period massive binaries in the young massive star forming region M17: }
\subtitle{Evidence for a large orbital separation at birth?}
   \author{H. Sana
          \inst{1}
          \and
          M. Ram\'irez-Tannus\inst{2}
          \and
          A. de Koter\inst{1, 2}
          \and
          L. Kaper\inst{2}
          \and
          F. Tramper\inst{3}
          \and
          A. Bik\inst{4}
   }

   \institute{Institute of astrophysics, KU Leuven, Celestijnlaan 200D, 3001 Leuven, Belgium\\
              \email{hugues.sana@kuleuven.be}
         \and
         Anton Pannekoek Institute for Astronomy, University of Amsterdam,
              Science Park 904, 1098 XH Amsterdam, The Netherlands\\
              \email{m.c.ramireztannus@uva.nl}
        \and
		 European Space Astronomy Centre (ESAC), 
		 Camino bajo del Castillo, s/n Urbanizacion Villafranca del Castillo, Villanueva de la Canada, E-28 692 Madrid, Spain
		\and
		Department of Astronomy, Stockholm University, 
		Oskar Klein Center, SE-106 91 Stockholm, Sweden      
             }

   \date{Received September 15, 1996; accepted March 16, 1997}

 
  \abstract
   {} 
   {The formation of massive stars remains poorly understood and little  is known about their birth multiplicity properties. Here, we aim to quantitatively investigate the strikingly low radial-velocity dispersion measured for a sample of 11 massive pre- and near-main-sequence stars (\srv\ $=5.6\pm0.2$~\kms) in  the  very young  massive star forming region M17, in order to obtain first constraints on the multiplicity properties of young massive stellar objects. }
   {We compute the radial-velocity dispersion of synthetic populations of massive stars for various multiplicity properties and we compare the obtained \srv\ distributions to the observed value. We specifically investigate two scenarios: a low binary fraction and a dearth of short-period binary systems.}
   {Simulated populations with low binary fractions ($f_\mathrm{bin}=0.12_{-0.09}^{+0.16}$) or with truncated period distributions ($P_\mathrm{cutoff}>9$~months) are able to  reproduce the low \srv\ observed within their 68\%-confidence intervals. Furthermore, parent populations with $f_\mathrm{bin}>0.42$ or $P_\mathrm{cutoff}<47$~d can be rejected at the 5\%-significance level. Both constraints are in stark contrast with the high binary fraction  and plethora of short-period systems in few Myr-old, well characterized  OB-type populations. To explain the difference in the context of the first scenario would require a variation of the outcome of the massive star formation process. In the context of the second scenario, compact binaries must form later on, and the cut-off period may be related to physical length-scales representative of the bloated pre-main-sequence stellar radii or of their accretion disks.}
   {If the obtained constraints for the M17's massive-star population are representative of  the multiplicity properties of massive young stellar objects, our results may provide support to a massive star formation process in which binaries are initially formed at larger separations, then  harden or migrate to produce the typical (untruncated) power-law period distribution observed in few Myr-old OB binaries.
}

   \keywords{binaries: spectroscopic --- Stars: early-type --- Stars: formation
     ---  open clusters and associations: individual: M17
               }

   \maketitle
%

\section{Introduction}\label{s:intro}

In the quest to obtain observational constraints on the poorly understood  massive star formation process \citep{ZiY07}, two lines of research are currently pursued. The first line  tries to get glimpses of the early phases of the formation process by penetrating the optically thick cloak of the emerging massive stars. 
This motivates observations at longer wavelengths and the search for the elusive massive star accretion disks. The second line of research attempts to constrain the formation process through some key properties of their end products. This includes constraints on the masses and rotation rates -- and their distributions -- as well as on their multiplicity properties. In this paper we focus on this latter aspect.

Spectroscopic searches for massive young stellar objects (mYSOs) in young ($<1$~Myr) star-forming regions were performed by, e.g., \citet{Bik06, Bik12}, \citet{Ochsendorf11} and \citet{Ellerbroek13}, but these analyses mostly focused on single star properties. The first search for young massive binaries was performed by \citet{ABK07}, who did a 2- to 3-epoch radial-velocity (RV) study of 16 embedded young massive stars in seven massive star forming regions. The authors identified two OB stars with RV variations of approximately 90~\kms\ and  measured a RV dispersion (\srv) of 35~\kms\ for the whole sample, and of 25~\kms\ after excluding the two close binaries. According to the authors, it was not possible to further statistically distinguish  their sample \srv\ from that of a single-star population given their large RV measurement errors.

 With $L=3.6\times10^6$~\Lsun\ \citep{Povich07} 
 and  an age of less than $1$~Myr \citep[submitted, hereafter \citetalias{Maria16}]{Hanson97, Broos07,Hoff08, PCB09,Maria16},  the giant H~{\sc ii} region \object{M17} is one of the most luminous star-forming regions in our Galaxy. 
 In \citetalias{Maria16}, we characterized nine candidate mYSOs and three OB stars, all with stellar masses in the range of 6 to 20~\msun.  Stars in this mass range dominate some of the samples from which massive star multiplicity statistics are derived  \citep[e.g.,][]{SdMdK12,KK14,DDS15} and are thus particularly relevant. Despite its modest size, this sample is one of the largest among  very young (likely $<$~1~Myr) clusters where massive stars can be caught just after their formation phase.

  \citetalias{Maria16}  revealed a lack of double-lined spectroscopic binaries and a narrow range of measured RVs ($-10 < v_\mathrm{rad} <20$~\kms). This is in stark contrast to the overall properties of small and large samples of fully formed, main-sequence massive stars \citep{2007A&A...474...77K,SdMdK12,SdKdM13,KiK12,KK14,DDS15,AST16}. These works have indeed established that 40\%\ to 50\%\ of OB-type binaries have a period of one month or less, with large RV amplitudes ($\Delta v_\mathrm{rad}>100$~\kms). Here we further investigate  the modest RV dispersion of the massive star population in M17 and  make a first attempt at quantifying its multiplicity properties.

\section{Observational constraints}\label{s:obs}
\subsection{Data overview}\label{s:data_overview}

  Our  M17's  sample is composed by 9 known candidate mYSOs\footnote{We implicitly assumed in the present work that mYSO tracers used to identify most of our sample stars are not affected by binarity.}  \citep{Hanson97,CEN80}  -- six of which were confirmed PMS stars in \citetalias{Maria16} --, and three are OB stars  observed in \citetalias{Maria16} to trace the early-type main sequence (B111, B164 and B253,  see \citetalias{Maria16} for a discussion on B253). 
 The known NGC6618's central binary CEN 1a (O4~V) as well as the CEN 18 and 37 binary candidates were not included because they were not mYSO candidates, nor were the remaining $\sim$40 other O and early B stars in the region  \citep{Hoff08}.

The data acquisition and data reduction are described in \citetalias{Maria16}. In short,
we acquired  optical to near-infrared (300--2500~nm) spectra of each star using the \xsh\ spectrograph \citep{2011A26A...536A.105V}, 
yielding spectra with a resolving power $\lambda/\Delta \lambda$ of 3,300 to 11,300 depending on the spectral range and slit width used. For three objects, spectra were  taken at multiple epochs (Table~\ref{t:rvs}). The location of the sample stars in M17's field-of-view is displayed in figure~1 of \citetalias{Maria16}.

\subsection{Radial velocities and RV dispersion}\label{ss:RVs}

We used the RV measurement method described in \citet{SdKdM13} and successfully applied to various sets of \xsh\ data in, e.g., \citet{SvBT13} and \citet{TSF16}. It simultaneously adjusts all desired spectral lines (Table~\ref{t:lines}) and all observational epochs (Table~\ref{t:rvs}) for a given object, taking into account the error spectrum or the signal-to-noise  ratio. We have modified the method to be able to use a 
combination of Lorentz profiles (for H lines) and Gaussian profiles (for He and other metal lines). Prior to the fitting, the residual of the nebular contamination affecting the core of the H and of some \hea\ lines were clipped from the data. No reliable measurement could be obtained for the PMS star B163 due to a lack of suitable lines and poor quality data. The star is excluded in the following. The obtained fits of the line profiles have a reduced $\chi^2$ in the range 0.6 to 1.3, and yield uncertainties on the measured RVs of less than 3~\kms\ (Table~\ref{t:rvs}). Larger uncertainties would only strengthen our results.

\begin{table}
\centering
\setlength{\tabcolsep}{3pt}
\caption{
Sample stars and journal of the observations. Sources classified as PMS in \citetalias{Maria16} appear in bold. Sp.Types are from \citetalias{Maria16}.
}
\label{t:rvs}
\begin{tabular}{cccccc}
\hline
\hline
\multicolumn{2}{c}{Identifier}   &  Sp.Type  &    MJD$-$      & $v_\mathrm{rad}$   & d\tablefootmark{(3)} \\
B92\tablefootmark{(1)}& CEN(OI)\tablefootmark{(2)}   &           &   2\,450\,000 & (\kms)    &   (pc) \\
\hline
\object{B111} 	&	2 (337)	&   O4.5V	&	6490.116 & $  3.4\pm  0.8$	&	0.695	\\
{\bf \object{B163}}	&	\dots		&   kA5	&	5455.982 &    \dots			&	0.154	\\
\object{B164}   	&	25		&   O6V		&	6490.123 & $ -1.8\pm  2.3$	&	0.372	\\
\object{B215} 	    &	\dots	&   O9-B1V	&	6115.249 & $ 12.3\pm  1.0$	&	0.856	\\
{\bf \object{B243}} 	&	51		&   B8V		&	6114.188 & $ 19.5\pm  1.8$	&	0.607	\\
\dots	    &	\dots	&   \dots	&	6490.140 & $ 20.2\pm  1.5$	&	\dots	\\
\object{B253} 	    &	26		&   B3-B5III &	6490.197 & $  9.5\pm  0.8$	&	0.573	\\
{\bf \object{B268}} 	&	49		&   B9-A0	&	6114.236 & $  1.4\pm  1.0$	&	0.686	\\
\dots      	&	\dots	&   \dots	&	6114.259 & $ 17.1\pm  1.1$	&	\dots	\\
\dots   	&	\dots	&   \dots	&	6490.169 & $  4.3\pm  1.3$	&	\dots	\\
{\bf \object{B275}} 	&	24		&   B7III	&	5054.125 & $-11.2\pm  1.4$	&	0.698	\\
\object{B289}    	&	31		&   O9.7V	&	5456.024 & $ -2.0\pm  2.5$	&	1.390	\\
\dots   	&	\dots	&   \dots	&	6114.290 & $ -2.3\pm  2.0$	&	\dots	\\
\object{B311}    	& 16 (258)	&   O8.5Vz	&	6490.097 & $  4.2\pm  0.4$	&	1.601	\\
{\bf \object{B331}} 	&	92		&   late-B	&	6115.276 & $ 13.9\pm  1.5$	&	1.194	\\
{\bf \object{B337}} 	&	93		&   late-B	&	6489.185 & $  7.5\pm  0.6$	&	1.171	\\
\hline
\srv & \dots    & \dots & \dots & $5.6\pm0.2$ & \dots \\
\hline
\end{tabular}
   \tablefoot{
\tablefoottext{1}{\citet{1992MsT..........2B}.} 
\tablefoottext{2}{Alternative IDs: CEN \citep{CEN80} and OI \citep{1976PASJ...28...35O}.}
\tablefoottext{3}{Projected distance of the stars to the center of the cluster ($\alpha = 275.124574$\degr,  $\delta = -16.179031$\degr).}
}
\end{table}

The obtained measurements  yield a 
RV dispersion of $\sigma_\mathrm{1D} = 5.6\pm0.2$~\kms\ (Table~\ref{t:rvs}). Restricting the computation of $\sigma_\mathrm{1D}$ to the candidate mYSOs or to the confirmed PMS stars in our sample (Table~\ref{t:rvs}) yield similar values of $5.9\pm0.2$  and $6.9\pm0.4$~\kms, respectively. These small differences have no impact on our results and we pursue our analysis with the full sample.  Among the three stars with two or three epochs, only B268 presents significant RV variations and is a promising binary candidate.

The modest \srv\ value that is observed is not expected if M17's mYSOs contains short-period binaries such as those that are frequently found in young stellar populations \citep[e.g.,][]{SaE11}.  These short-period binaries are  easily detected in spectroscopy  and usually dominate the measured RV dispersion.  Indeed, other studies of OB star populations with only a few epochs all measured a much larger velocity dispersion of the order of 25 to 35~\kms\ \citep[e.g.,][]{BSM01,ABK07,HBES12}.

\section{Multiplicity properties}\label{s:mult}

In this section, we investigate two possible scenarios to explain the observed low RV dispersion
of the young massive star population in M17: {\it (i)} a lower binary fraction, or {\it (ii)} a dearth of short-period binaries. We ignore the role of eccentricity and mass-ratio as these have a smaller impact on \srv.

\subsection{Methodology}\label{ss:simul}
To quantify the multiplicity properties of massive stars in M17, we compare the RV dispersion  of the observed sample with that resulting from Monte Carlo population synthesis computed with different underlying multiplicity properties. Each parent population, hence each Monte-Carlo run, is characterized by a primary- and a single-star mass function, a given
binary fraction (\fbin), a set of orbital parameter distributions and a cluster velocity dispersion ($\sigma_\mathrm{cl}$). 

We draw  the masses of the single  and of the primary stars from a Kroupa mass function ranging from  $6$ to $20$~\msun. We  adopt the multiplicity properties derived by 
\citet{SdKdM13} for Galactic young open clusters  as the baseline in our population synthesis. These include $f_\mathrm{bin}=0.7$ and $f_{P_\mathrm{orb}}\propto (\log_{10} P_\mathrm{orb})^{-0.5}$, with $0.15 \le \log_{10} (P_\mathrm{orb}/\mathrm{day}) \le 3.5$ and a flat mass-ratio distribution.  Adopting an \"Opik law all the way down to orbital periods of 1~day  would  not change our conclusions.

The cluster velocity dispersion ($\sigma_\mathrm{cl}$) is computed from the virial theorem, assuming a total mass for the M17 cluster of $1.4 \times 10^3$~\Msun\ \citep{StG90} and a typical 1~pc cluster radius. This yields $\sigma_\mathrm{cl}=2$~\kms. Other published mass estimates  range from a few times $10^3$ to $6\times10^4$~\msun\ \citep{PBSO15,PCB09}, corresponding to $\sigma_\mathrm{cl}=1.4$~to 5.2~\kms. Adopting the former value has no impact on our results while picking the latter one would only strengthen our conclusions because it leaves almost no room for binary contributions to the observed \srv\ value. In the following, we adopt $\sigma_\mathrm{cl}=2$~\kms.

Each Monte Carlo run is formed by $10^5$ realisations and uses the sample size, number of epochs, time sampling and measurement accuracy of the observational sample. The density distributions of the simulated \srv\  for given parent multiplicity properties are then constructed and used for tests of hypotheses that allow us to accept or reject the specific parent population at various significance levels.
\subsection{A low binary fraction}\label{s:fbin}

\begin{figure}[t!]
  \includegraphics[width=\columnwidth]{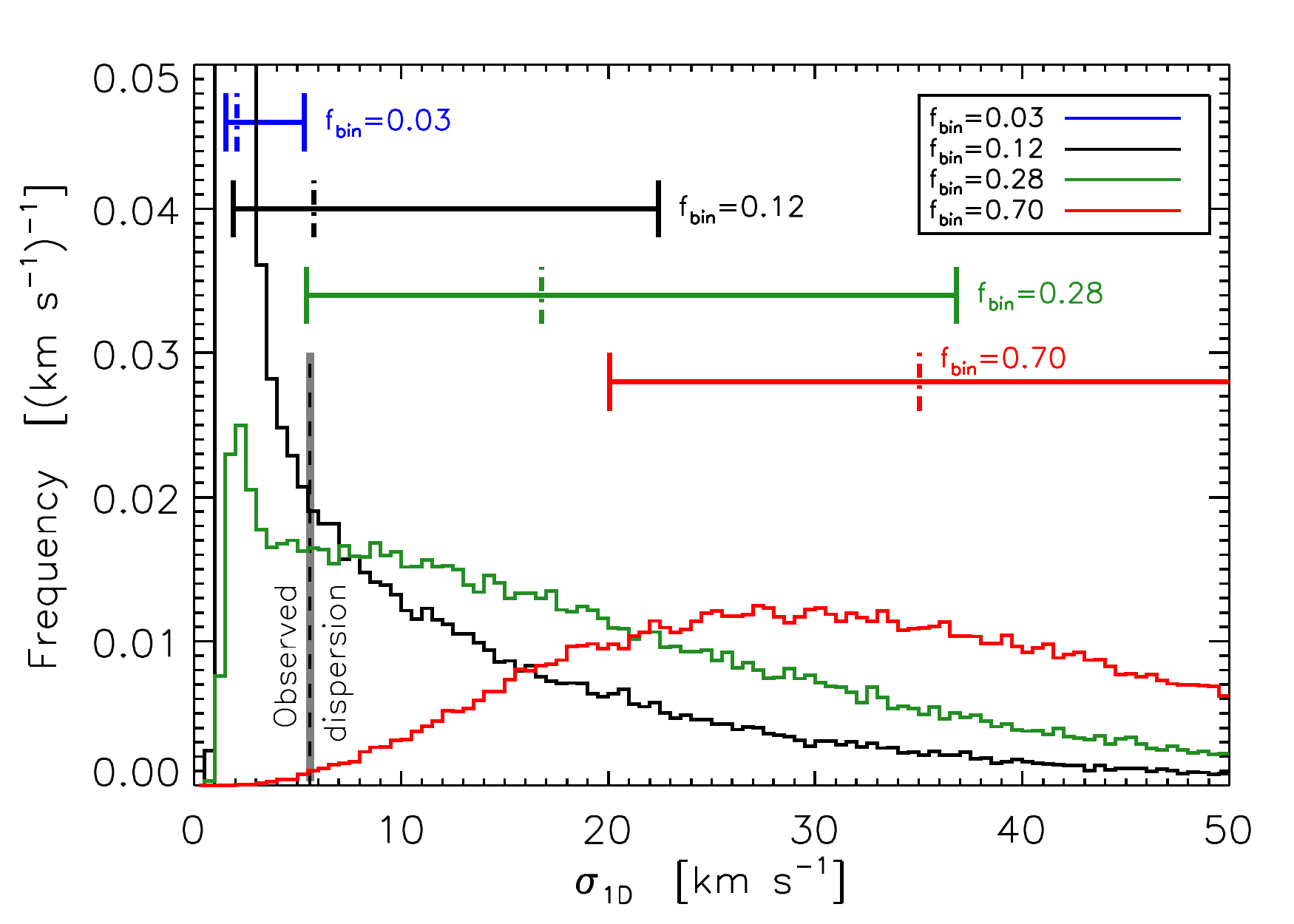}
  \caption{Simulated \srv\ distributions for different parent binary fractions. The vertical dashed line indicates the observed \srv\ for our M17 sample stars. The median, 0.16 and 0.84 percentiles of the simulated distributions are indicated on the upper part of the graph. The distribution corresponding to $f_\mathrm{bin}=0.03$ is not shown for clarity; it is almost entirely dominated by the velocity dispersion of the cluster and thus peaks strongly at 2~\kms\ ($\approx \sigma_\mathrm{cl}$).}
  \label{f:fbin}
\end{figure}

We first investigate the possibility that the low RV dispersion observed in M17 results from a low binary fraction. In our MC experiment, we keep the distribution of the orbital properties fixed  
and we vary the binary fraction from 0.00 to 1.00 in steps of 0.01. The obtained \srv\ distributions for representative values of \fbin\ are displayed in Fig.~\ref{f:fbin}.

The distribution whose median corresponds to our observed \srv\ has $f_\mathrm{bin}=0.12$. Distributions with $f_\mathrm{bin}=0.03$ to $0.28$ can all match the $\sigma_\mathrm{1D}=5.6$~\kms\ within their 0.16 to 0.84 percentiles, that is, their predictions agree within $\pm 1 \sigma$ with the observed value. We thus conclude that, given standard orbital parameter distributions, the observed velocity dispersion in M17 is best reproduced by a low binary fraction $f_\mathrm{bin}=0.12^{+0.16}_{-0.09}$.

Similarly, no simulation with $f_\mathrm{bin}\geq 0.34$ (resp. 0.42) can reproduce M17's \srv\ within their 0.90 (resp. 0.95) percentiles and the corresponding hypotheses can thus be rejected at the 10 and 5\%-significance levels, respectively.

\subsection{A truncated period distribution}\label{s:pcrit}

We now assume that the binary fraction is consistent with constraints from OB stars in young open clusters, that is, $f_\mathrm{bin}=0.7$ for $P_\mathrm{orb}<3500$~d \citep{SdMdK12}, but that the binary population is composed of longer-period binaries, which typically results in a lower \srv\ \citep{GSPZ10,HBES12}. In practice, we adopt a truncated period distribution where the orbital periods of all binaries with $P_\mathrm{orb}<P_\mathrm{cutoff}$ are iteratively re-drawn from the parent distribution until they are equal to or larger than $P_\mathrm{cutoff}$.

\begin{figure}[t!]
  \includegraphics[width=\columnwidth]{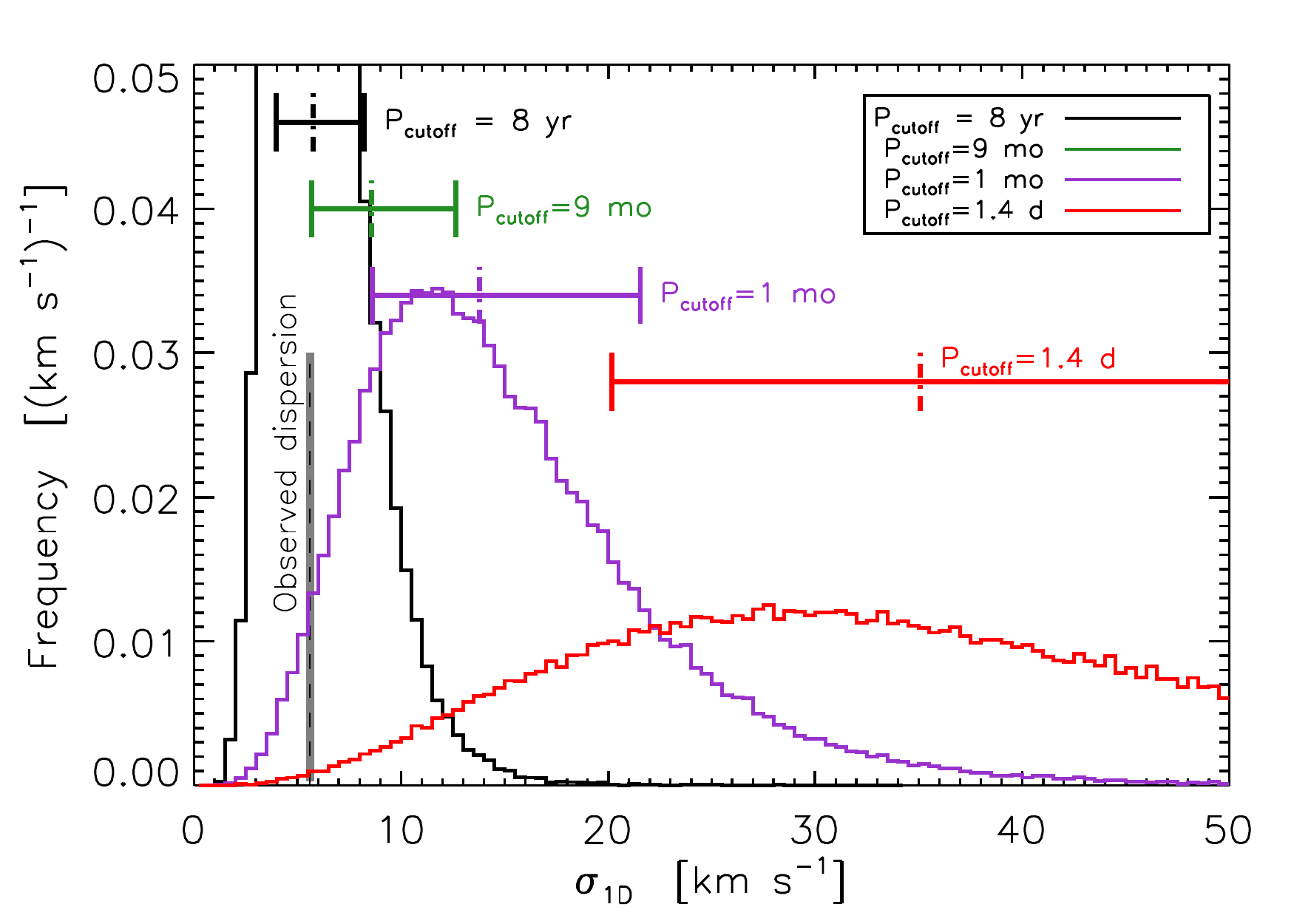}
  \caption{
  Same as Fig.~\ref{f:fbin} for different cutoff periods ($P_\mathrm{cutoff}$).  
  The distribution corresponding to $P_\mathrm{cutoff}\approx9$~mo is intermediate between those of $P_\mathrm{cutoff}\approx8$~yr and 1~month and is not shown for clarity.} 
  \label{f:pcrit}
\end{figure}

Resulting distributions for representative values of $P_\mathrm{cutoff}$ are shown in Fig.~\ref{f:pcrit}. Distributions with  \pcutoff\ $\approx$ 8~yr have a median that best matches the  M17's observed \srv\ while distributions with $P_\mathrm{cutoff} \gtrsim 283$~d ($\approx 9$~months) are all able to reproduce the observed \srv\ within their 68\%\ confidence range. Parent populations with $P_\mathrm{cutoff}<131$~d (resp. 47~d) can be rejected at the 10 and 5\%-significance levels, respectively.

  \section{Discussion}\label{s:discuss}

  \subsection{A low binary fraction?}

  Under the assumption that the  distribution of orbital parameters in M17 is not different from that of other well characterized OB-star populations, a low RV dispersion can only be explained by a low binary fraction. This, however, would make M17 quite atypical.  Populations of massive stars in clusters across the entire mass range have all revealed a large binary fraction, from modest star-forming regions such as the Orion Nebula and Sco-Cen \citep{PBH99, 2007A&A...474...77K}
  to massive associations with several tens of thousands of \msun\ of stellar content, such as, Cyg~OB2 \citep{KK14}, the Carina nebula \citep{SaE11} and the Tarantula region in the Large Magellanic Cloud \citep{SdMdK12, DDS15}.

  Nearby young open clusters such as IC\,1805, NGC\,6231, M16, and NGC~6611 have a similar mass to that of M17. With typical ages of  several Myr, they could be viewed as M17's older siblings. Yet, they  all have a detected binary fraction of over 40\%\ before any detection bias correction \citep{DBRM06,SGN08,SGE09,SJG11}. This would invalidate any temporal sequence, suggesting a variation in the outcome of the star-formation process (unless massive stars are paired at a later stage).

  An alternative scenario that would produce a low binary fraction but that does not invoke variations in the star-formation process may involve runaways stars, which  are indeed preferentially 
     single objects \citep{MHG09,SLBL14}. Given that the region is too young to have produced any supernovae, 
     runaways in M17 should result from dynamical ejection. This would imply that {\it (i)}   M17 has already gone through core collapse, {\it (ii)}   the runaway ejection velocity is low (to preserve a low \srv), and {\it (iii)}   the initial massive star binary fraction was either low or lacked short-period systems. Short-period (hard) binaries are indeed expected to be ejected almost undisturbed and should still be seen in the runaway population if they were present in  M17's original population.

  The last possibility is that our sample is dominated by binary interaction products, which are known to  most likely appear as single stars in RV studies \citep{dMSL14}. Again, the young age of the region challenges this explanation unless these stars have exchanged mass or merged well within their pre-main-sequence evolution. The latter scenario cannot be widespread  as it would be incompatible with the significant population of short-period binaries detected in the young  clusters mentioned earlier.
 
  \subsection{A lack of short-period systems?}

  The existence of many short-period binaries is a well documented observational fact \citep{GCM80,MGH98,SaE11}. Direct measurements of the orbital period distribution  further confirm the relative abundance of short-period binaries, with at least 20\%\ of all OB-type binaries having a period of less than a week; and 40\%\ to 50\%, a period of less than a month \citep{SdMdK12,KK14, DDS15,AST16}.
  
    However, the process that leads to the formation of these short-period systems  remains  mysterious. Core fission has fallen into demise \citep{Toh02} while other formation mechanisms such as disk fragmentation \citep{KKMK09,KMK10} and dynamical capture \citep{FPZ11} predict binaries with typical orbital periods that are two to four orders of magnitude too long.

  The possible lack of short-period binaries in one of the youngest populations of massive stars ever characterized is thus particularly intriguing and may suggest that massive binaries are originally formed at  larger typical separations. The newly born binary systems should then harden on a time-scale of the order of 1~Myr or less to match the observational properties of main-sequence massive star populations of only a few Myr of age.

  In the context of disk fragmentation theories, such an inward migration process may be driven by the interaction with the remnants of the accretion disk or with other, likely smaller mass, proto-stellar bodies. Migration may then  stop with  the disappearance of the sink of angular momentum, either as the circumbinary disk is destroyed or as other proto-stellar  bodies are pushed far out or are ejected.

  It is noteworthy that the minimum period cutoff that we derived ($P_\mathrm{cutoff} =47$~d) corresponds to the typical size of a  (bloated) massive pre-main-sequence object ($R\sim100$~\Rsun) while our most likely cut-off period ($P_\mathrm{cutoff} \approx 8$~yr) corresponds to approximately 10~AU in a 10~\msun\ system, that is, similar to the size of the first direct detection of the remnant of an accretion disk around a massive star \citep{KHM10}. While more and better constraints are desirable, it seems possible that the truncation of the period distribution, if real, can be related to meaningful physical scale-lengths in the formation process.

  \section{Conclusions}\label{s:ccl}

  We have investigated the low radial-velocity dispersion ($\sigma_\mathrm{1D}=5.6\pm0.2$~\kms) found in a sample of 11 massive pre/near main-sequence stars in the M17 region. With an age of less than 1~Myr, this region is one of the youngest massive star forming regions for which a quantitative investigation of the multiplicity properties has been performed. We show that the low RV dispersion found is incompatible with multiplicity properties derived from older populations of massive stars.

The present results from the M17 region seem to be corroborated by the multiplicity properties of Tr14, a dense young cluster \citep[$<1$~Myr,][]{SMG10} in the Carina nebula. Though it has only six O stars with RV information \citep{SaE11}, none of them has been detected as a short-period binary \citep[though see][for HD~93129A]{MASB16}. 

If the multiplicity properties of the present M17's sample (and that  of Tr14) were found to be representative of an earlier phase in the formation process than that of currently well characterized OB-star regions -- as  suggested by the younger age of M17 -- our results would support a formation mechanism where binaries are initially born at larger separations (100~\rsun\ or more). 
They would also support the presence of a migration process that would harden the systems on a time-scale of the order of 1~Myr, or less, in order to match the observed multiplicity properties of OB-type populations.  

While these speculations would fit well in a refinement of  disk fragmentation theories, higher-quality observational constraints (larger samples, more observational epochs) are needed. Theoretical and/or numerical computations to investigate, among others, the conditions for such a migration process to work are also highly desirable. These may lead to new insight into the origin of the relative universality of the  period distribution of OB stars, which is an input ingredient of population synthesis needed to investigate the frequency of gravitational wave events.

\begin{appendix}
\section{Online table}
\begin{table*}
\centering
\caption{Spectral lines available for RV measurements.} \label{t:lines}
\begin{tabular}{lccccccccccc}
\hline
\hline
Object  & B111 &B164&B215&B243&B253&B268&B275&B289&B311&B331&B337\\
\hline
\hea\,\l4026 & x &    & x  &    & x  &    & x  &    & x  &    &      \\
\hea\,\l4120 &   &    &    &    &    &    &    &    & x  &    &     \\
\hea\,\l4143 &   &    &    &    &    &    &    &    & x  &    &     \\
\heb\,\l4200 & x & x  &    &    &    &    &    &    &    &    &     \\
\hea\,\l4387 &   &    & x  &    & x  &    & x  &    & x  &    &     \\
\hea\,\l4471 & x & x  &    &    & x  &    & x  & x  & x  &    &     \\
\mgb\,\l4481 &   &    &    &    &    &    &    &    & x  &    &     \\
\heb\,\l4541 & x & x  &    &    &    &    &    &    &    &    &     \\
\heb\,\l4686 & x &    &    &    &    &    &    & x  & x  &    &     \\
\hea\,\l4713 &   & x  &    &    &    &    &    &    & x  &    &     \\
\hea\,\l4922 &   & x  &    &    &    &    &    & x  & x  &    &     \\
\hea\,\l5015 &   & x  &    &    &    & x  & x  &    & x  &    &     \\
\hea\,\l5048 &   &    &    &    &    &    &    &    & x  &    &     \\
\heb\,\l5412 & x & x  &    &    &    &    &    & x  & x  &    &     \\
\hea\,\l5876 & x &    & x  &    &    &    &    &    & x  &    &     \\
\hgam      & x & x  & x  & x  & x  & x  & x  & x  & x  &    &     \\
\hbet      & x & x  &    & x  & x  & x  & x  & x  & x  &    &     \\
\halp      &   & x  & x  &    & x  &    &    & x  & x  &    &     \\
Pa-21      &   &    &    &    &    &    &    &    &    &    & x   \\
Pa-15      &   &    &    &    &    & x  & x  &    &    & x  &     \\
Pa-14      &   &    &    &    & x  & x  & x  &    &    &    &     \\
Pa-13      &   &    &    &    & x  & x  & x  &    &    &    &     \\
Pa-12      & x & x  & x  & x  & x  & x  & x  & x  & x  & x  & x   \\
Pa-11      & x & x  & x  & x  & x  & x  & x  & x  & x  & x  & x   \\
\peps      &   &    &    &    &    &    &    &    &    &    & x   \\
\pdelt     &   &    &    &    & x  &    &    &    &    &    &     \\
\pgam      & x &    &    &    &    &    &    & x  & x  &    &     \\
\pbeta     & x & x  &    &    & x  &    &    & x  & x  &    &     \\
\fea\,\l8621 &   &    &    &    & x  &    & x  &    &    & x  & x    \\
\sid\,\l4088 &   &    &    &    &    &    &    &    & x  &    &     \\
\sid\,\l4116 &   &    &    &    &    &    &    &    & x  &    &     \\
\nc\,\l13176 &   &    & x  &    & x  & x  &    &    &    &    &    \\ 
\hline
\end{tabular}
\end{table*}

\end{appendix}
\end{document}